\documentclass[a4paper,11pt]{article}
\usepackage[latin1]{inputenc}
\usepackage[swedish,english]{babel}
\usepackage{a4wide}
\usepackage{authblk}
\usepackage{amsmath,amsthm}
\usepackage{amssymb}
\usepackage{algorithm}
\usepackage{algorithmic}
\usepackage{ifpdf}
\ifpdf
  \usepackage[pdftex]{graphicx}
  \usepackage{epstopdf}
\else
  \usepackage[dvips]{graphicx}
\fi
\usepackage{subcaption}
\usepackage{sidecap}
\usepackage[pdftitle={Scalable modeling of cell populations in continuous time},
  pdfauthor={Baker, Engblom, Wilson},
  pdffitwindow=true,
  breaklinks=true,
  colorlinks=true,
  urlcolor=blue,
  linkcolor=red,
  citecolor=red,
  anchorcolor=red]{hyperref}
\usepackage[numbers,sort&compress]{natbib}

\numberwithin{equation}{section}
\numberwithin{table}{section}
\numberwithin{figure}{section}

\theoremstyle{plain}

\theoremstyle{definition}
\newtheorem{definition}{Definition}[section]
\newtheorem{assumption}[definition]{Assumption}

\theoremstyle{remark}

\newcommand{\Nvoxels}{N_{\mbox{{\tiny vox}}}}
\newcommand{\Ncells}{N_{\mbox{{\tiny cells}}}}

\newcommand{\Omegaext}{\Omega_{\mbox{{\tiny ext}}}}

\newcommand{\rhoprol}{\rho_{\mbox{{\tiny prol}}}}
\newcommand{\rhodeath}{\rho_{\mbox{{\tiny death}}}}
\newcommand{\rhodeg}{\rho_{\mbox{{\tiny deg}}}}
\newcommand{\kappaprol}{\kappa_{\mbox{{\tiny prol}}}}
\newcommand{\kappadeath}{\kappa_{\mbox{{\tiny death}}}}
\newcommand{\Iadhesion}{I_{\mbox{{\tiny adhesion}}}}

\newcommand{\review}[1]{#1}

\title{Scalable population-level modeling of biological cells
  incorporating mechanics and kinetics in continuous time}

\author[1]{Stefan Engblom\thanks{Corresponding author: S.~Engblom,
    telephone +46-18-471 27 54, fax +46-18-51 19 25.}}
\author[2]{Daniel B. Wilson}
\author[2]{Ruth E. Baker}

\affil[1]{{\footnotesize Division of Scientific Computing, Department
    of Information Technology, Uppsala University, SE-751 05 Uppsala,
    Sweden. E-mail:
    \href{mailto:stefane@it.uu.se}{stefane@it.uu.se}.}}

\affil[2]{{\footnotesize Wolfson Centre for Mathematical Biology,
    Mathematical Institute, University of Oxford, Radcliffe
    Observatory Quarter, Oxford, OX2 6GG, United Kingdom. E-mail:
    \href{mailto:daniel.wilson@maths.ox.ac.uk}{daniel.wilson@maths.ox.ac.uk},
    \href{mailto:ruth.baker@maths.ox.ac.uk}{ruth.baker@maths.ox.ac.uk}.}}

\date{}

\begin{document}

\selectlanguage{english}

\maketitle

\begin{abstract}

  The processes taking place inside the living cell are now understood
  to the point where predictive computational models can be used to
  gain detailed understanding of important biological phenomena. A key
  challenge is to extrapolate this detailed knowledge of the
  individual cell to be able to explain at the population level how
  cells interact and respond with each other and their environment. In
  particular, the goal is to understand how organisms develop,
  maintain and repair functional tissues and organs.

  In this paper we propose a novel computational framework for
  modeling populations of interacting cells. Our framework
  incorporates mechanistic, constitutive descriptions of biomechanical
  properties of the cell population, and uses a coarse graining
  approach to derive individual rate laws that enable propagation of
  the population through time. Thanks to its multiscale nature, the
  resulting simulation algorithm is extremely scalable and highly
  efficient. As highlighted in our computational examples, the
  framework is also very flexible and may straightforwardly be coupled
  with continuous-time descriptions of biochemical signalling within,
  and between, individual cells.
 









\end{abstract}


\section{Introduction}
\label{sec:intro}

Development, disease and repair all require the tightly coordinated
action of populations of cells. In almost every case a plethora of
interacting components, acting on a range of spatial and temporal
scales, combines to drive the observed tissue-level
behaviours. Research in the biosciences is now advanced to the stage
where we have sufficient understanding of intercellular processes to
build relatively sophisticated models of a wide range of cellular
behaviours. A promising approach to generate, test and refine
hypotheses as to the relative contributions of various mechanisms to
tissue-level behaviours is that of cell-based computational modeling:
individual cells are explicitly represented, each cell has a position
that updates over time, and cells may also have an internal state or
program determining their behaviour. Cell-based models hold great
potential in this regard because they can naturally capture both
stochastic effects and cell-cell heterogeneity, and they can be used
to explore tissue-level behaviours when complex hypotheses on the
cellular scale prevent straightforward continuum approximations at the
tissue level. \review{Recent applications of cell-based models to
  study population-level behaviours include embryonic
  development~\cite{Farhadifar:2007:TIO,Mao:2011:PPA,
    Trichas:2012:MCR,Monier:2010:AAB,Atwell:2015:MLM}, wound
  healing~\cite{Walker:2004:ABC,Vermolen:2013:ASC,Ziraldo:2013:CMO}
  and tumour
  growth~\cite{Anderson:1998:CAD,Alarcon:2003:CAM,Ramis-Conde:2008:MTI,
    Rejniak:2007:AIB,Naumov:2011:CAM}.}

Multiple cell-based modeling approaches exist, and they can be
categorised according to their approaches to representing cell
positions as being either on- and off-lattice. In the on-lattice
approach, space is divided up into a discrete grid of lattice sites. A
common type of on-lattice model is the cellular automaton, in which
each cell occupies a single lattice site and attempts to move to a new
site at each time step according to a set of update rules. This volume
exclusion rule can be relaxed to allow multiple cells per lattice
site, depending on the level of spatial description required by the
problem under consideration. Position update rules typically take into
account the number and type of neighbouring cells, and can also depend
on other information associated with lattice sites, such as nutrient
or signalling factor concentrations~\cite{Hatzikirou:2010:LGC}. Hybrid
models often use systems of ordinary differential equations (ODEs) or
partial differential equation (PDEs) to model the evolution of
biochemical concentrations (see, for
example,~\cite{Alarcon:2003:CAM,Ramis-Conde:2008:MTI,Robertson:2015:IOM,
  Starrus:2014:MAU}). Position update rules can also be stochastic, so
that cells move according to, e.g., a biased random
walk~\cite{Othmer:1997:ABC, Othmer:1988:MDB,
  Benazeraf:2010:RCM,Yates:2010:RSP,Setty:2012:AMO,
  Thompson:2012:MCM}. \review{In most cases a regular lattice is used,
  for example square or hexagonal, but unstructured lattices have also
  been used in several cases
  (see~\cite{Kansal:2000:SBT,Radszuweit:2009:CTG} and references
  therein).} A different on-lattice technique is the cellular Potts
model, wherein each cell is allowed to occupy multiple lattice sites,
and energy minimisation is used to propagate the shape of each cell
over time. \review{The cellular Potts model has been used to study
  biological processes ranging from cell
  sorting~\cite{Graner:1992:SBC,Zhang:2011:CSC} and
  morphogenesis~\cite{Vasiev:2010:MGC,Hester:2011:MCM} to tumour
  growth~\cite{Szabo:2013:CPM,Scianna:2012:HCP}, and methods to
  provide a macroscopic limit to these models have been
  provided~\cite{Alber:2007:CML,Lushnikov:2008:MDB}.}

\review{A basic off-lattice approach to cell-based modeling is the
  cell-centre-based model which assumes cells are, in effect, point
  particles that interact with each other via some specified potential
  function~\cite{Drasdo:2005:ASC,Meineke:2001:CMO,Ramis-Conde:2008:MTI}.}
Meanwhile, in vertex
models~\cite{Kursawe:2017:IOI,Weliky:1990:MBC,Fletcher:2014:VMO} cell
populations are modelled as a tessellation of polygons or polyhedra,
whose vertices move due to forces originating from the cells, whilst
in immersed boundary
models~\cite{Cooper:2017:NAO,Rejniak:2007:AIB,Tanaka:2015:SFF,
  Tanaka:2015:LBI} cell boundaries are represented as a set of points
that move like elastic membranes immersed in a fluid. Other cell-based
models take the subcellular composition of cells into account, such as
the subcellular element model~\cite{Newman:2007:SCB} or the finite
element vertex model~\cite{Brodland:1994:ETM}.

\review{Each cell-based modeling approach has specific advantages and
  disadvantages, and incorporate different representations of both
  biochemical signalling and
  biomechanics~\cite{Drasdo:2007:OTR,Starrus:2014:MAU,Tanaka:2015:SFF,
    Hoehme:2010:ACB,Van-Liedekerke:2015:STM,Osborne:2017:CIB}.} For
example, an immersed boundary model allows a detailed representation
of cell shapes, but this benefit comes at an increased computational
cost in comparison to other methods. Cellular Potts models are very
versatile in the possible effects that can be modelled but, because
updates are controlled via a Metropolis-type algorithm together with a
user-specified Hamiltonian, time can only be measured in terms of
Monte Carlo steps. When modeling a specific application, it is
necessary to weigh the benefits of the existing cell-based models
against each other in the context of the specific
application. \review{A comprehensive review of mechanocellular models,
  including a table that compares the different modelling approaches,
  is provided
  in~\cite{Drasdo:2007:OTR,Van-Liedekerke:2015:STM,Fletcher:2017:MCM}
  and a comparison of different model output is detailed
  in~\cite{Van-Liedekerke:2015:STM,Osborne:2017:CIB}.}

\review{In this work we will focus on lattice-based approaches that
  allow a user-specified maximum number of cells to occupy each
  lattice
  site~\cite{Kansal:2000:SBT,Dormann:2002,Hatzikirou:2010:LGC}.} On a
given structured or unstructured tessellation of space we develop
constitutive equations governing the dynamics of the cell
population. As in the cellular Potts model, our update rules are
stochastic and are established from global calculations. An important
difference, however, is that our simulations take place in continuous
time, thus allowing for a meaningful coupling to other continuous-time
models. Our rationale for developing this approach is a desire to be
able to quickly and efficiently simulate three-dimensional tissues
consisting of large numbers of cells, with inclusion of biomechanical
effects, intercellular signalling and external inputs. Since our
approach is to base the modeling of cell biomechanics on the Laplace
operator over the spatial tessellation, we refer to our method as
\emph{Discrete Laplacian Cell Mechanics} (DLCM). Relying on the
discrete Laplace operator is advantageous because one may make use of
highly developed and scalable numerical methods to evolve the
biomechanical details of the cell population over time. An important
advantage provided by our highly efficient framework is then the
potential ability to conduct parameter sensitivity analysis, parameter
inference and model selection for cell-based models; these are
increasingly important research tools in this era of quantitative,
interdisciplinary biology.

The outline of this work is as follows: in \S\ref{sec:methods} we
detail the model and its computational implementation; in
\S\ref{sec:results} we describe four examples that showcase the
utility of our modeling framework; and in \S\ref{sec:discussion} we
summarise our results and discuss avenues for future exploration.


\section{Methods}
\label{sec:methods}

The method we propose is developed by distributing the cells onto a
grid of \emph{voxels} and defining a suitable physics over this
discrete space. The Laplace operator emerges as a convenient and basic
choice to describe evolution of the biomechanics of the population,
but more involved alternatives could also be employed in its
place. \review{We enforce a bound on the number of cells per voxel
  such that processes at the scale of individual cells may be
  meaningfully described on a voxel-local basis. For the simulations
  performed in this paper the voxels contain a maximum of
  two cells, but larger carrying capacities than this can also be
  supported. The choice of discretisation (and so the maximum number
  of cells that can be accommodated in any voxel) should be made on a
  case-by-case basis, taking into account the need to balance
  computational complexity with the extent to which data on
  individual-cell-level processes are available.} By evolving the
individual cells via discrete PDE operators, e.g. the discrete
Laplacian, processes at the population level are connected in an
efficient and scalable way to those taking place inside the individual
cells. In \S\ref{subsec:overview} we offer an intuitive algorithmic
description of our framework, and a more formal development is found
in \S\ref{subsec:formal}.

\subsection{Informal overview of the modeling framework}
\label{subsec:overview}

\begin{figure}[htp]
  \centering
  \begin{subfigure}{0.4\textwidth}
    \includegraphics[clip=true,trim=8cm 10cm 8cm 10cm]{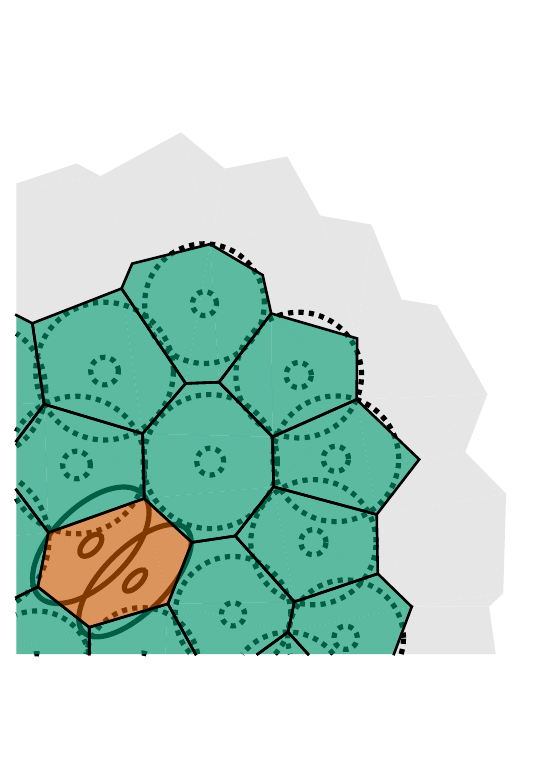}
    \caption{}
    \label{fig:basic_model:a}
  \end{subfigure}
  \begin{subfigure}{0.4\textwidth}
    \includegraphics[clip=true,trim=8cm 10cm 8cm 10cm]{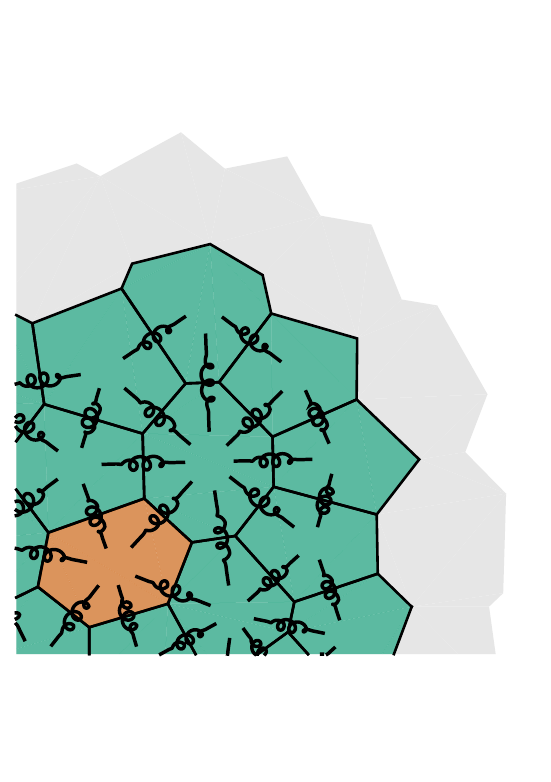}
    \caption{}
    \label{fig:basic_model:b}
  \end{subfigure}
  \caption{Schematic explanation of the numerical model. An
    unstructured Voronoi tessellation (\ref{fig:basic_model:a}) with
    \textit{green} voxels containing single cells and a \textit{red}
    voxel containing two cells. The modeling physics for the cellular
    pressure can be thought of as if the pressure was spread evenly
    via linear springs connecting the voxel centres
    (\ref{fig:basic_model:b}). The grid here is unstructured, but the
    same derivation is used for regular, e.g.~Cartesian and hexagonal,
    grids.}
  \label{fig:basic_model}
\end{figure}

We consider a computational grid consisting of voxels $v_i$, $i =
1,\ldots,\Nvoxels$, in two or three spatial dimensions. We make no
assumptions as to whether the grid is structured or not, however, we
do require that a consistent Laplace operator may be formed over the
grid. Each voxel $v_i$ shares an edge with a neighbour set $N_i$ of
other voxels. In two dimensions, each voxel in a Cartesian grid has
four neighbours and on a regular hexagonal lattice, each voxel has six
neighbours. On a general unstructured triangulation, each vertex of
the grid has a varying number of neighbour vertices and, in this
general and flexible case, the voxels themselves can be constructed as
the polygonal compartments of the corresponding dual Voronoi diagram
(Figure~\ref{fig:basic_model}).

At any given point in time the voxels are either empty or may contain
a certain number of cells. If the number of cells is at or below
\emph{the carrying capacity}, the system is assumed to be at
steady-state. Hence in the absence of any other processes, the cell
population is then completely static. If the number of cells in one or
more voxels exceeds the carrying capacity, the cells push each other
and exert a cellular pressure. Eventually, this non-equilibrium state
is changed by an event, for example one of the cells moves into a
neighbouring voxel, and the pressure is redistributed. This goes on
until, possibly, the system relaxes into a steady-state.

What is the relevant constitutive equation for this cellular pressure?
We make a more detailed investigation of this in \S\ref{subsec:formal}
below, but chiefly, for an isotropic medium and a scalar potential,
thus essentially assuming the pressure to be spread evenly as in
Figure~\ref{fig:basic_model:b}, the answer is that the pressure is
distributed according to the negative Laplacian, with source terms for
all voxels where the carrying capacity is exceeded.

At any instant in time, a pressure gradient between two neighbouring
voxels induces a force that, in turn, can cause the cells within the
voxels to move. The rate of this movement is proportional to the
pressure gradient, with a conversion factor that may depend on the
nature of the associated movement. For example, it may be reasonable
to assume that a cell may move into an empty voxel or into an already
occupied voxel with different rates per unit of pressure gradient.

The simulation method is event-based and takes the form of an outer
loop over successive events, see~Algorithm~\ref{alg:main}, lines
\ref{main:ln:while}--\ref{main:ln:endwhile}. Because the dynamics is
driven by a discrete numerical PDE operator, i.e.~the Laplacian in our
context, we call the simulation method \textit{Discrete Laplacian Cell
  Mechanics} (DLCM). It should be noted that it might be beneficial in
certain situations to use a different driving PDE operator: for a
concrete example, see \S\ref{subsec:gradient}.

For any given state of the cell population, the cellular pressure is
calculated and the rates of all possible events are determined
(lines~\ref{main:ln:pressure_rates}--\ref{main:ln:other_rates}). Here
the sampling procedure of Gillespie \cite{Gillespie:1976:GMN} may be
used; the sum of all the rates decides the time for the next event,
and a proportional sampling next determines the event that happens
(lines~\ref{main:ln:Gillespie:1976:GMN1}--\ref{main:ln:Gillespie:1976:GMN3}).
Until the time of this next event, any other processes local to each
voxel may be simulated in an independent fashion
(line~\ref{main:ln:other}). Finally, as the event is processed, a new
cell population state is obtained and the loop starts anew.

\begin{algorithm}[htb!]
\caption{Outline of the DLCM simulation methodology.}
\label{alg:main}
\begin{algorithmic}[1]

  \STATE{\textit{Initialise:} at time $t = 0$, given a state $(u_i)$,
    $u_i \in \{0,1,2\}$ over (a subset of) the mesh of voxels $(v_i)$,
    $i = 1,\ldots,\Nvoxels$.}
  \label{main:ln:init}

  \WHILE{$t < T$}
  \label{main:ln:while}

  \STATE{Solve for the cellular pressure $(p_i)$,
    eqs.~\eqref{eq:dlaplace}--\eqref{eq:hom_dirichlet}. Compute all
    movement rates $(r_j)$ for the subset of voxels where the cells
    may move, eqs.~\eqref{eq:D1}--\eqref{eq:D3}.}
  \label{main:ln:pressure_rates}

  \STATE{Determine also the rates for any other processes taking place
    in the model such as proliferation and death events, or active
    migration.}
  \label{main:ln:other_rates}

  \STATE{Compute the sum $\lambda$ of all transition rates thus
    computed.}
  \label{main:ln:Gillespie:1976:GMN1}

  \STATE{Sample the next event time by $\tau = -\log(U_1)/\lambda$,
    for $U_1$ a uniformly distributed random variable in $(0,1)$.}

  \STATE{Determine which event happened by inverse sampling: find $n$
    such that
    $\sum_{j=1}^{n-1} r_j < U_2 \lambda \le \sum_{j=1}^n r_j$, for
    $U_2$ a second $U(0,1)$-distributed random number.}
  \label{main:ln:Gillespie:1976:GMN3}

  \STATE{Update the state of all cells with respect to any other
    continuous-time processes taking place in $[t,t+\tau)$, e.g.,
      intracellular kinetics or cell-to-cell communication.}
    \label{main:ln:other}

  \STATE{Update the state $(u_i)$ by executing the state transition
    associated with the determined event.}

  \STATE{Set $t = t + \tau$.}

  \ENDWHILE \label{main:ln:endwhile}

  \STATE{\textit{Result:} a sampled outcome of the system consisting
    of states observed at discrete times $(t_k) \in [0,T]$.}
\end{algorithmic}
\end{algorithm}

In summary, our new modelling framework is lattice-based, allows for a
range of physical and biomechanical phenomena to be accurately
described, and facilitates explicit modelling of cell size/excluded
volume effects through the pressure-dependence of cell movements. The
incorporation of stochasticity via the Gillespie algorithm renders it
naturally able to represent the noise observed in most biological
systems. Moreover, it is simple to simultaneously integrate either
deterministic or stochastic models of biochemical reaction
networks. Through the choice of sensible numerical linear algebra
techniques, the framework is efficient and even complex models can be
simulated in short times. This means that it will be possible to
perform statistical inference of model parameters using quantitative
experimental data together with techniques such as approximate
Bayesian computation~\cite{Toni:2009:ABC,Jagiella:2017:PAH}.

\subsection{Formal description}
\label{subsec:formal}


We now develop the details of the DLCM framework. At some instant in
time, let the grid $(v_i)$, $i = 1,\ldots,\Nvoxels$, be populated with
$\Ncells$ cells. In the interest of a transparent presentation we only
allow each voxel to be populated with $u_i \in \{0,1,2\}$ cells; other
arrangements may also be useful, but a suitable physics for voxels
populated \emph{below} the carrying capacity should then depend on
biological details such as the tendency of the cells to stay in close
proximity to each other.

Due to the spatial discretisation and the discrete counting of cells,
the task is to track changes over this chosen state space. In
continuous time this amounts to figuring out which cell will move to
what voxel, and when it will move. This requires a governing physics
defined over the discrete state. A continuous-time Markov chain
respects the ``memoryless'' Markov property and stand out as a
promising approach, requiring only movement \emph{rates} in order to
be fully defined.  Our model of the population of cells follows from
three equations \eqref{eq:continuity}, \eqref{eq:Darcy}, and
\eqref{eq:heat} below, understood and simplified under three
assumptions, Assumptions~\ref{ass:1}--\ref{ass:3}. We present each in
turn as follows.

Let $u = u(t,x)$ represent the cell density at time $t$ and at the
point $x$. We need to keep in mind that our numerical model is to be
formulated on an existing grid of voxels $(v_i)$ containing a bounded
(integer) number of cells $u_i$. Hence the continuum limit $h \to 0$
(of some suitable measure of the voxel size going to zero) is not
meaningful. However, we shall use a continuous notation initially for
ease of presentation. Our starting point is the continuity equation
\begin{align}
  \label{eq:continuity}
  \frac{\partial u}{\partial t} + \nabla \cdot I &= 0,
\end{align}
where $I$ is the current, or flux. Since we are aiming at an
event-based simulation we will later use eq.~\eqref{eq:continuity} to
derive rates for discrete events in a continuous-time Markov chain. To
prescribe the current $I$ we now make a starting assumption.

\begin{assumption}
  \label{ass:1}
  The tissue is in mechanical equilibrium when all cells are
  placed in a voxel of their own.
\end{assumption}

Assumption~\ref{ass:1} expresses the idea that small Brownian-type
movements of each cell about its (voxel-) center can be ignored. It
does not exclude an additional description of any \emph{active}
movements, such as chemotaxis or haptotaxis. With sufficient
conditions for equilibrium specified, it follows from
Assumption~\ref{ass:1} that only doubly occupied voxels will give rise
to a rate to move, and we will describe this increased rate as a
pressure source. In the absence of any other units we can set this
pressure source to unity identically.

Let $p = p(t,x)$ denote the cellular pressure at time $t$ and position
$x$, again using a continuous notation for variables which will be
implemented on a discrete grid. Interpreting the current $I$ as the
result of a pressure gradient, we take the simple phenomenological
model
\begin{align}
  \label{eq:Darcy}
  I &= -D\nabla p,
\end{align}
which, with the interpretation $D \equiv \kappa/\mu$, the quotient
between the permeability $\kappa$ and the viscosity $\mu$, is a form
of Darcy's law in porous fluid flow. Notably, Darcy's law can be
obtained from first principles using homogenisation arguments
\cite{DarcyMS}. Our use of eq.~\eqref{eq:Darcy}, however, will be
different as cells preserve their geometrical integrity and do not
flow freely like a fluid. Rather, the current in eq.~\eqref{eq:Darcy}
will be understood as a rate of the discrete event that a cell changes
voxel.

To complete the model we require a constitutive equation relating $u$
and $p$. With the cellular pressure driven by sources in the form of
overcrowded voxels, we provide a constitutive model of pressure
evolution using the heat equation,
\begin{align}
  \label{eq:heat}
  \varepsilon \frac{\partial p}{\partial t} &= \Delta p+s(u),
\end{align}
with $s(u)$ a source function that will be prescribed below. Specifically,
eq.~\eqref{eq:heat} follows by assuming an isotropic medium and a scalar
potential pressure. For each voxel populated at or below the carrying
capacity, there is no net flux of the potential and the divergence
theorem implies the Laplace operator. For an overpopulated voxel,
there is instead a net outward flux, then captured via the divergence
theorem as a source term.

\medskip

Eq.~\eqref{eq:heat} is a time-dependent PDE and would be complicated
to handle within the current context. To move forward we therefore
need to bring in an additional assumption.

\begin{assumption}
  \label{ass:2}
  The cellular pressure of the tissue relaxes rapidly to equilibrium
  in comparison with any other mechanical processes of the system.
\end{assumption}

In cases where the biochemical kinetics of the individual cells also
affect their mechanical behaviour, for example via signal transduction
or proliferation, Assumption~\ref{ass:2} entails that these processes
must occur on a slower time-scale than the propagation of the cellular
pressure. Importantly, Assumption~\ref{ass:2} simplifies
eq.~\eqref{eq:heat} into the non-singular $\varepsilon \to 0$ limit,
the Laplacian equilibrium,
\begin{align}
  \label{eq:laplace}
  -\Delta p  &= s(u). \\
  \intertext{The most immediate boundary conditions from
  eq.~\eqref{eq:heat} are}
  p \bigr|_{\partial \Omega_D} &= 0 \mbox{ (free boundary), and, } \\
  (\partial p/\partial n)\bigr|_{\partial \Omega_N} &= 0
 \mbox{ (solid wall)}.
\end{align}
The domain $\Omega$ understood here generally consists of the bounded
subset of $\mathbb{R}^2$ or $\mathbb{R}^3$ which is populated by the
cells. Its boundary $\partial \Omega$ can be written as
$\partial \Omega = \partial \Omega_D \cup \partial \Omega_N$, the
Dirichlet and the Neumann boundaries, respectively, which can be
chosen according to the specific biological problem under
consideration. It is also possible to interpolate between the two
boundary conditions,
\begin{align}
  \left[ \alpha p+(\partial p/\partial n) \right] \bigr|_{\partial
  \Omega_R} &= 0 \mbox{ (semi-free boundary)},
\end{align}
to model, for example, an increasingly impenetrable cellular matrix as
$\alpha \to 0$ by a homogeneous Robin condition. To simplify the
presentation here, we employ homogeneous Dirichlet conditions
($\partial \Omega = \partial \Omega_D$) throughout the computational
examples in \S\ref{sec:results}.

\begin{SCfigure}
  \includegraphics[clip=true,trim=8cm 10cm 8cm 10cm]{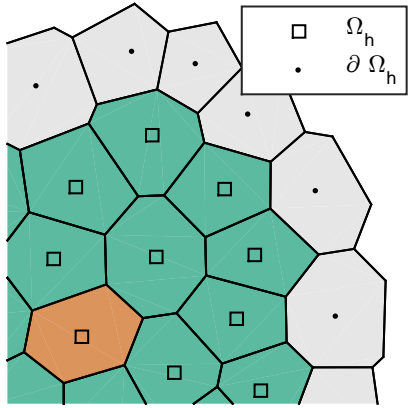}
  \caption{The definition of the discrete numerical domain $\Omega_h$,
    the set of populated voxels, and the numerical boundary $\partial
    \Omega_h$, the set of unpopulated voxels sharing an edge with
    $\Omega_h$. \vspace{1.25cm}}
  \label{fig:basic_model_h}
\end{SCfigure}

We now re-interpret the developed model onto the given tessellation
$(v_i)$, aiming specifically for a time-continuous and event-driven
simulation. Denote by $\Omega_h$ the subset of voxels $v_i$ for which
$u_i \not = 0$. Similarly, let $\partial \Omega_h$ denote the discrete
boundary; this is the set of unpopulated voxels that are connected
(i.e.~share an edge) with a voxel in $\Omega_h$. See
Figure~\ref{fig:basic_model_h} for an illustration. We first consider
the discrete version of eq.~\eqref{eq:laplace},
\begin{align}
  \label{eq:dlaplace}
  -L p  &= s(u), \; i \in \Omega_h, \\
  \label{eq:hom_dirichlet}
  p_i &= 0, \; i \in \partial \Omega_h,
\end{align}
where the source term is given by, in non-dimensionalized form,
$s(u_i) = 0$ for $u_i \le 1$ and $s(u_i) = 1$ whenever $u_i = 2$, in
view of the normalisation $p = 0$ for the static case, following
Assumption~\ref{ass:1}. In eq.~\eqref{eq:dlaplace}, $L$ is a discrete
Laplacian over the currently active grid $\Omega_h$. The precise
choice of (consistent) numerical method chosen to define $L$ is not
likely to have a strong influence on the model output since the
Laplacian operator is quite forgiving to such details. In our
experiments we used a finite-element-based discrete Laplacian operator
with linear basis functions and a lumped mass-matrix, which simplifies
to traditional finite difference stencils on structured grids. It
follows that the continuous interpretation of the source term is
formally consistent with a Dirac delta function; in continuous
language, $s(u) = s(u(x))$ is just $\sum_{u_i >c} \delta_{v_i}(x)$,
where $v_i$ here denotes the center point of the $i$th voxel and $c$
the carrying capacity.

We now go back to considering the induced movement of the cells
according to the continuity equation \eqref{eq:continuity}. By
eq.~\eqref{eq:Darcy}, a pressure gradient gives rise to a proportional
current; $I \propto -\nabla p$. Let us write $I(i \to j)$ to denote
the current from voxel $v_i$ to neighbour voxel $v_j$. In our discrete
setting this current is calculated by integrating the pressure
gradient across the edge between the two voxels,
\begin{align}
  \label{eq:current}
  I(i \to j) \propto -\int_{v_i \cap v_j} \nabla p(x)
  \, \text{d}\vec{S} = \frac{e_{ij}}{d_{ij}} (p_i-p_j) =: I_{ij},
\end{align}
with $d_{ij}$ the distance between voxel centres, $e_{ij}$ the shared
edge length, and where the simplification takes the discrete nature of
the model into account by simply substituting a scaled difference for
the gradient. In eq.~\eqref{eq:current}, the current is reversed $(j
\to i)$ when the result is negative. It follows that, in the presence
of a pressure source, a non-zero pressure gradient results everywhere
except on solid (Neumann) boundaries. It remains to specify $D$ in
eq.~\eqref{eq:Darcy}, including the subset of cells that are free to
move as a result of this pressure gradient.

\begin{assumption}
  \label{ass:3}
  The cells in a voxel occupied with $n$ cells may only move into
  a neighbouring voxel if it is occupied with less than $n$
  cells.
\end{assumption}

Let us write $R(e)$ for the rate of the event $e$ and use the notation
$i \to j$ for the particular event that one cell moves from voxel $i$
to $j$. Under our present framework, $u_i$ is constrained to
$\{0,1,2\}$ and a total of three distinct cases emerges for the
movement rates:
\begin{alignat}{3}
  \label{eq:D1}
  &R(i \to j;\;\mbox{$u_i \ge 1$, $u_j = 0$, $v_j$ never visited}) &= 
  D_1 I_{ij}; \\
  \label{eq:D2}
  &R(i \to j;\;\mbox{$u_i \ge 1$, $u_j = 0$, $v_j$ visited previously}) \; &= 
  D_2 I_{ij}; \\
 \label{eq:D3}
  &R(i \to j;\;\mbox{$u_i > 1$, $u_j = 1$}) &= 
  D_3 I_{ij}.
\end{alignat}
Here, $D_1$ is the conversion factor from a unit pressure gradient
into a movement rate to a voxel $v_j$ never visited before, $D_2$
similarly, but into a voxel previously occupied, and $D_3$ covers the
``crowding'' case where a cell in a doubly occupied voxel enters a
singly occupied voxel.

\begin{SCfigure}
  \centering
  \includegraphics[width = 0.4\textwidth,clip=true,trim=0.5cm 0 0 0]{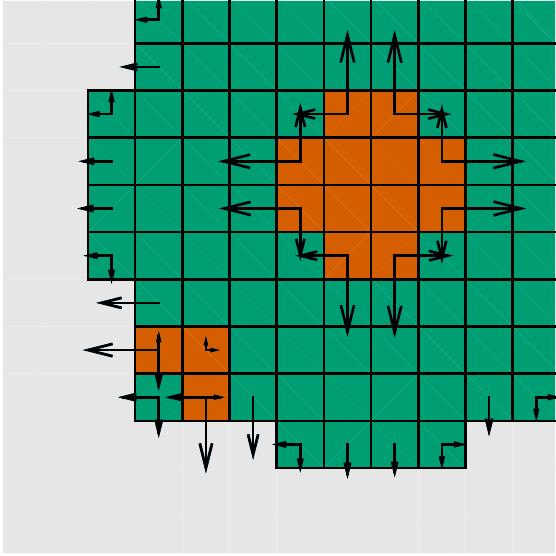}
  \caption{Schematic illustration of the model: as in
    Figure~\ref{fig:basic_model}, \textit{green} voxels contain single
    cells and \textit{red} voxels contain two cells, giving rise to
    pressure sources. This pressure is propagated by a Laplacian
    operator over the voxels and induces a force, and hence a rate to
    move, for the cells in the subset of voxels indicated by the
    arrows. Cells in boundary voxels may move into empty voxels
    and cells in doubly populated voxels may move into voxels
    containing fewer cells. All arrows are drawn on a common
    scale.}
  \label{fig:basic_schematics}
\end{SCfigure}

Although dependent on the specific application under consideration, we
generally expect to have $D_1 \ll D_2$, since the extracellular matrix
contained in previously unvisited voxels is usually thought to be less
penetrable than that of previously occupied voxels. To understand the
scale of $D_3$ we note that from the divergence theorem
\begin{align}
  -\int_{\partial \Omega} (\nabla p \cdot n) \, \text{d}\vec{S} &= 
  \int_{\Omega} s(u) \, \text{d}V.
\end{align}
In other words, the total pressure gradient over any closed surface
$\partial \Omega$ is equal to the enclosed sources. With $\partial
\Omega$ the boundary surrounding the whole cell population, it follows
that the ratio $D_2/D_3$ expresses the preference for events at the
tissue boundaries (cells entering empty voxels) to events internal to
the region (cells in doubly occupied voxels moving to an already
occupied neighbouring voxel). For example, assuming the presence of a
single internal source voxel and that $D_2/D_3 \equiv 1$, then the
probability of a cell in the source voxel to move is the same as the
total probability of a cell to move into any of the boundary voxels in
$\partial \Omega$. The rates in eqs.~\eqref{eq:D1}--\eqref{eq:D3} are
illustrated on a common scale with $D_1 = D_2 = D_3$ in
Figure~\ref{fig:basic_schematics}. Note that, on this Cartesian grid,
by the integration over an edge in eq.~\eqref{eq:current}, the current
is non-zero only along the cardinal directions up/down and right/left,
respectively.

\review{In addition to the rates in eqs.~\eqref{eq:D1}--\eqref{eq:D3},
  one can easily extend the basic framework to include active cell
  movements and/or behaviours, as prescribed via additional rates and
  to be handled in
  lines~\ref{main:ln:pressure_rates}--\ref{main:ln:other_rates} of
  Algorithm~\ref{alg:main}.
For example, a possible approach to include the effects of cell-cell
adhesion is to define the resistance force for a cell moving from
voxel $i$ to $j$ due to cell-cell adhesion with cells in the non-empty
voxel $k$ as
\begin{align}
  \label{eq:adhesion}
  a_{ijk} &:= \alpha e_{ik} 
  \min(0,\vec{r}_{ij} \cdot \vec{r}_{ik}),
\end{align}
for some adhesion constant $\alpha$ and for neighbor voxel pairs
$(i,j)$ and $(i,k)$. Here $e_{ik}$ is the edge length over which
adhesion bonds are active and $(\vec{r}_{ij},\vec{r}_{ik})$ denote
unit vectors pointing from voxel $i$ to $j$ and, respectively, from
voxel $i$ to $k$. The $\min$-construct in eq.~\eqref{eq:adhesion}
ensures that adhesion acts as a resistance (i.e.~negative)
force. Summing up the adhesion forces due to all populated neighbor
voxels of voxel $i$ and scaling by the inter-voxel distance $d_{ij}$
we thus evaluate the net adhesion current contribution as
\begin{align}
  \delta \Iadhesion(i \to j) &= d_{ij}\sum_{k \in N_i} a_{ijk},
\end{align}
which is to be added to the current in eq.~\eqref{eq:current}. The
resistance force works against the pressure-driven current in
eq.~\eqref{eq:current} to reduce cell movements.}

Other options to capture similar effects could be to modify the
pressure source function and/or the boundary conditions; the best
choice is clearly dependent on the details of the modeling situation.


\section{Results}
\label{sec:results}

We now present some computational results for the proposed modeling
framework. In doing so we focus on highlighting the three distinct
advantages of our approach: (1) the mechanics is based on constitutive
equations obeying a well-understood physics; (2) the framework is fast
and scalable, and therefore convenient to experiment with; and (3), it
is also flexible and allows for a seamless coupling to other processes
of interest.

Throughout the examples below we constructed our computational grid
over a rectangular base geometry; we used structured Cartesian and
hexagonal grids in 2D and in one case a Cartesian grid in 3D. The grid
was initially populated by cells in the middle of the domain,
cf.~line~\ref{main:ln:init} of Algorithm~\ref{alg:main}. As previously
mentioned, we relied upon a finite-element-based discrete Laplacian
operator defined using linear basis functions and a lumped
mass-matrix, implemented in the Matlab add-on package \textit{PDE
  Toolbox} \cite{PDE}. The discrete Laplacian was inverted using
Matlab's built-in LU-decomposition (line~\ref{main:ln:pressure_rates}
of Algorithm~\ref{alg:main}); this was fast enough for our purposes
and in all tests performed here. Larger models may well benefit from
more advanced linear solver techniques. All code and the required data
is available, and we refer the reader to
\S\ref{subsec:reproducibility} for details.

In \S\ref{subsec:artefacts} we look at the possible dependency of
results on the underlying computational grid, and in
\S\ref{subsec:gradient} we investigate how to couple the cellular
behaviour with signals in the local external environment. In
\S\ref{subsec:delta_notch} we couple the modeling framework with
intracellular continuous-time processes, including cell-to-cell
signalling and, finally, in \S\ref{subsec:dynamics} we demonstrate
how the complex behaviours of a tumour model may be investigated.

\subsection{Simulated cell population is free of grid artefacts}
\label{subsec:artefacts}

\begin{figure}[htp]
  \centering
  \begin{tabular}{ccc}
    \includegraphics[width = 0.25\textwidth,clip=true,trim=1cm 0.5cm 1cm 0.5cm]{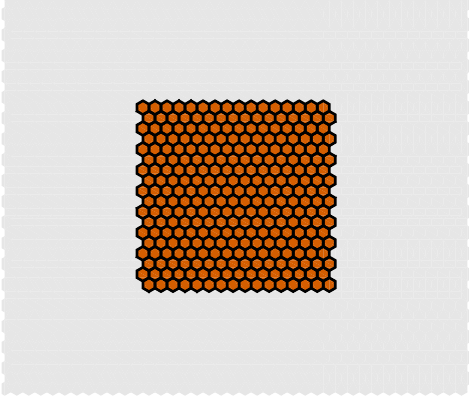}&
    \includegraphics[width = 0.25\textwidth,clip=true,trim=1cm 0.5cm 1cm 0.5cm]{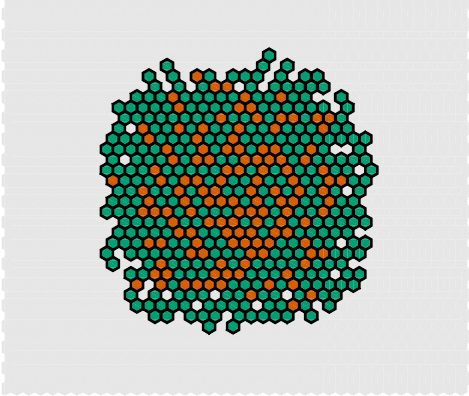}&
    \includegraphics[width = 0.25\textwidth,clip=true,trim=1cm 0.5cm 1cm 0.5cm]{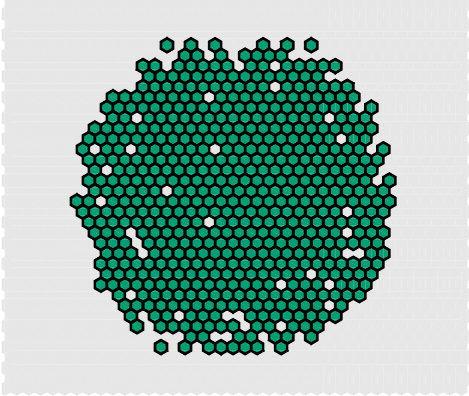}\\
    \includegraphics[width = 0.25\textwidth,clip=true,trim=0cm 0cm 0cm 0cm]{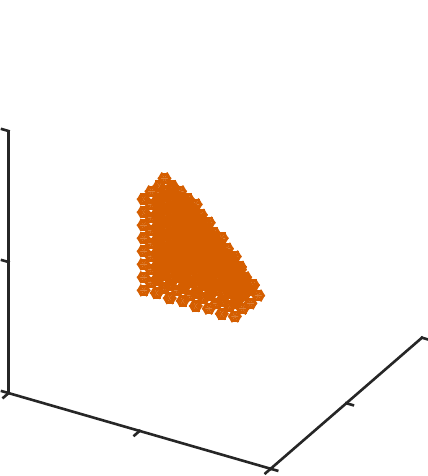}&
    \includegraphics[width = 0.25\textwidth,clip=true,trim=0cm 0cm 0cm 0cm]{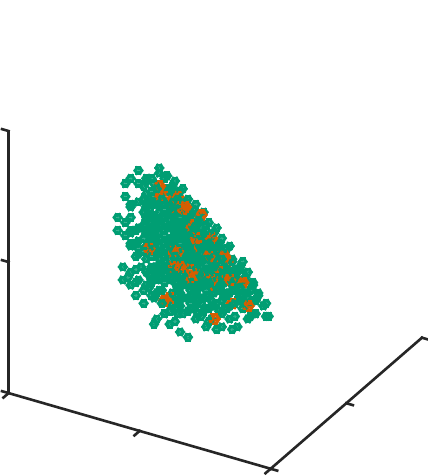}&
    \includegraphics[width = 0.25\textwidth,clip=true,trim=0cm 0cm 0cm 0cm]{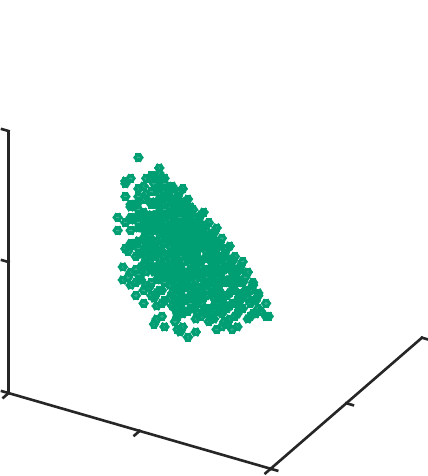}
  \end{tabular}
  \caption{The set-up of the relaxation process
    experiment. \textit{Top left:} all voxels in a square initially
    contain two cells (red voxels), \textit{top middle and right} (at
    2.5 and 100 units of time, respectively): after relaxation the
    populated domain becomes approximately circular in shape and all
    voxels contain a single cell (green voxels). The bottom row shows
    the equivalent experiment in three space dimensions over a
    Cartesian grid (at 0.5 and 20 units of time, respectively). For
    visibility only the voxels with coordinates $(x,y,z)$ satisfying
    $x+y+z \le 0$ are shown.}
  \label{fig:basic_test_samples}
\end{figure}

A disadvantage with grid-based methods and local update rules that has
been highlighted in the literature is that grid artefacts may appear
unless some care is exercised \cite{Yates:2013:IMF}. We test for
grid artefacts by artificially forcing a number of cells into a
square configuration and then relaxing the system to equilibrium
(Figure~\ref{fig:basic_test_samples}). In the absence of any other
mechanical forces, clearly, the equilibrium state should be
approximately circular.

\begin{figure}[htp]
  \centering
  \includegraphics[clip=true,trim=0cm 0cm 0cm 0cm]{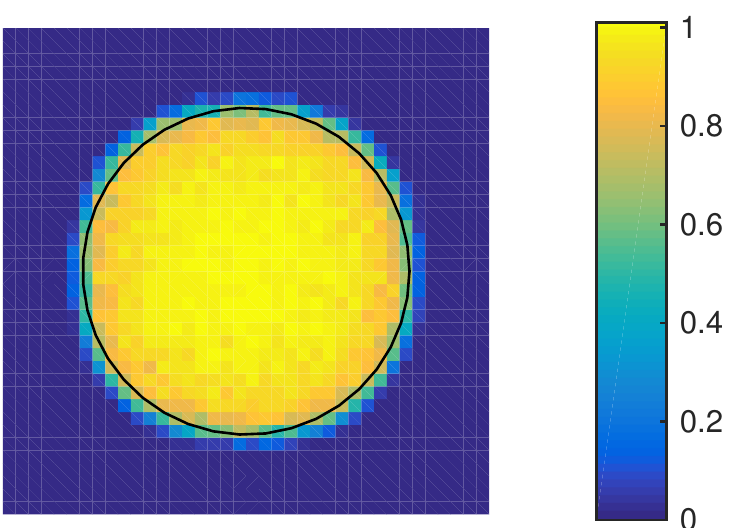}
  \includegraphics[clip=true,trim=0.35cm 0.15cm 0.35cm 0.15cm]{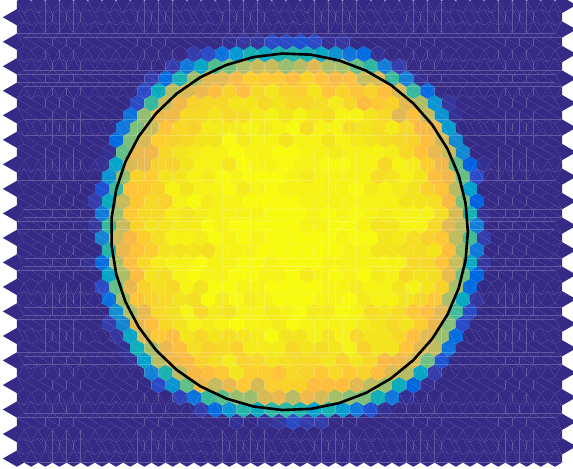}
  \caption{Average concentration ($N = 100$ trials) after relaxation
    from square initial data. \textit{Left:} Cartesian grid,
    \textit{right:} hexagonal grid. The reference circle has the same
    area as the total cell population. Although a discrete effect is
    visible, there are no grid artefacts in the sense of preferred
    expansion directions.}
  \label{fig:basic_test}
\end{figure}

In Figure~\ref{fig:basic_test} we quantify the average cellular
density across $N = 100$ independent runs of the model and find that,
although a very small memory effect from the initial data is still
visible (left/right and up/down), there is no dependence on the
underlying grid except for the natural distortions due to
discreteness. This comes from the fact that the physics is based on
well-understood constitutive equations and that we employ a
grid-consistent discretisation of the Laplacian
(eqs.~\eqref{eq:laplace} and \eqref{eq:dlaplace}).

\subsection{Population-level behaviour via sensatory control}
\label{subsec:gradient}

\begin{SCfigure}[1.3][h]
  \centering
  \includegraphics{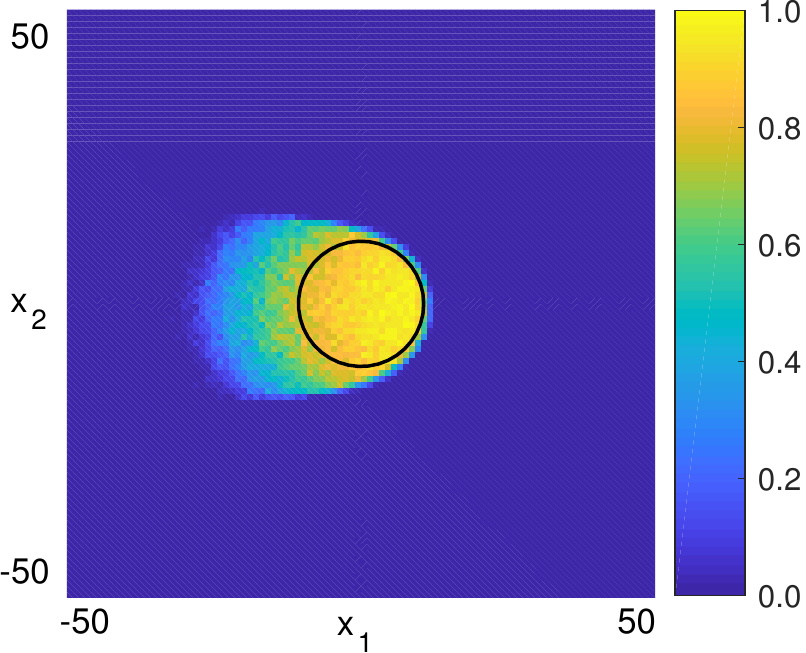}
  \caption{Average cell density ($N = 100$ trials) after relaxation
    from an initially circular explant (indicated by the black circle)
    with two cells in each voxel. The source of Slit is on the
    right-hand edge of the domain, at $x_1=X_s=50$, and we
    see that, on average, cells move preferentially to the left, away
    from the source of Slit.}
  \label{fig:gradient_growth}
\end{SCfigure}

A notable extension of this basic framework is to model a population
of cells that respond to external stimuli in their local
environment. An advantage of the DLCM framework is that we are not
restricted to considering a Laplacian operator but can freely choose
other PDE operators, depending on the mechanics of the problem under
consideration. As a test case we take the two-dimensional migratory
model of neurons responding to an inhibiting chemical known as
Slit~\cite{Cai:2006:MDG}. The cellular pressure, $p$, of the neurons
is now described by the chemotaxis equation
\begin{align}
  \label{eq:chemo_sensing}
  \varepsilon_p \dfrac{\partial p}{\partial t} & = \nabla \cdot \left(
  \nabla p + p \chi_1 \nabla S \right) + s(u),
\end{align}
where $\chi_1$ is the chemotactic sensitivity of the neurons to the
chemical $S$, i.e., the concentration of Slit. The parameter
$\varepsilon_p$ is taken to be small and represents our assumption
that the pressure equilibrates on a faster timescale than the movement
of the neurons. The source term, $s(u)$, is as defined
previously. Eq.~\eqref{eq:chemo_sensing} represents the movement of
cells down the gradients of the Slit concentration. We set the domain
for the experiment to be $\vec{x} = (x_1,x_2) \in \mathbb{R}^2$ and
the equation governing the concentration of Slit is given by
\begin{align}
  \varepsilon_S \dfrac{\partial S}{\partial t} &=
  D_S \Delta S - k S + Q \delta \left(x_1 - X_s \right), 
\end{align}
where $D_S$ is the constant diffusivity of Slit, $k$ is the rate of
degradation and $Q$ is the strength of the Slit source on the line
$x_1=X_s$. The parameter $\varepsilon_S$ is again taken to be small
and represents the assumption that the diffusion of Slit occurs on a
faster timescale than the movement of the cells. Imposing
Assumption~\ref{ass:2} thus allows us to solve the steady-state
problem
\begin{align}
  - \Delta p & = \nabla \left( p \chi_1 \nabla S \right) + s(u), \\
  \label{eq:slit_stst}
  0 & = D_S \Delta S - kS + Q \delta \left(x_1-X_s\right),
\end{align}
where the boundary conditions for the Slit chemical are
\begin{align}
  \label{eq:slit_bcs1} 
  S(x_1,x_2) \rightarrow 0, & \qquad \text{as } |x_1| \rightarrow \infty, \\
  \label{eq:slit_bcs2}
  \dfrac{\partial S}{\partial x_1} \rightarrow 0, &
  \qquad \text{as } |x_1| \rightarrow \infty.
\end{align}
We note that eqs.~\eqref{eq:slit_stst}--\eqref{eq:slit_bcs2} can be
solved analytically to give
\begin{align}
  S(x_1,x_2) = \dfrac{Q}{2\sqrt{k D_S}} e^{-\sqrt{k/D_S} |x_1 - X_s|}.
\end{align}
We also include a pressure-independent movement mechanism,
representing the active motion of the cells due to the chemo-repellent
(see Assumption~\ref{ass:1}). To include the chemotaxis-driven
movement of a cell from voxel $i$ to voxel $j$ we derive the following
current
\begin{align}
  \label{eq:current_chemo}
  I(i \to j) = - \chi_2 \int_{v_i \cap v_j} \nabla S(\vec{x})
  \, \text{d} \vec{\mathrm{S}} = \chi_2 \frac{e_{ij}}{d_{ij}} (S_i-S_j),
\end{align}
where $\chi_2$ is a measure of the affinity of the cells to Slit for
this type of pressure-independent movement. We initialise a circular
region, with radius $r=10$, of doubly occupied sites in the centre of
the domain, where the voxel size is taken to be $h=1$. We specify the
parameters $[\chi_1,\chi_2,D_S,k,Q,X_s] =
[100,5,50,0.1,2,50]$, and run $100$ simulations until terminal time $T
= 1000$. We present the average cell density profile in
Figure~\ref{fig:gradient_growth}. As expected, on average the cell
population moves away from the source of Slit, consistent with it
being a chemo-repellent.

\subsection{Continuous-time mechanics allows for seamless coupling to
  cellular signalling processes}
\label{subsec:delta_notch}

A great flexibility with the proposed framework comes from the fact
that additional processes may be simulated concurrently with the
overall mechanics of the cell population
(see~Algorithm~\ref{alg:main}, line~\ref{main:ln:other}). Notably,
such processes may include cell-to-cell communication driving cell
fate differentiation. To illustrate this we consider the classical
Delta-Notch intracellular signalling model~\cite{Collier:1996:PFL},
which produces pattern formation via a simple lateral inhibition
feedback loop. With $(n_i,d_i)$, respectively, the Notch and Delta
concentrations within cell $i$, this model takes the dimensionless
form
\begin{align}
  &\begin{array}{rcl}
    \label{eq:delta_notch_ODE}
    n'_i &=& f(\bar{d}_i)-n_i, \\
    d'_i &=& v \left( g(n_i)-d_i \right), \\
  \end{array}
  \intertext{where $'$ denotes differentiation with respect to time and}
  \label{eq:dave}
  \bar{d}_i &= \frac{1}{|N_i|} \sum_{j \in N_i} d_j, \qquad
  f(x) \equiv \frac{x^k}{a+x^k}, \qquad g(x) \equiv \frac{1}{1+bx^h}.
\end{align}
We take parameters $[a,b,v,k,h] = [0.01,100,1,2,2]$ and the average in
eq.~\eqref{eq:dave} is taken over the set of neighbours $N_i$ of the
cell $i$. Additional stochastic noise terms may also be added to
eq.~\eqref{eq:delta_notch_ODE}, but for simplicity we let the model be
fully deterministic.

To produce a dynamic population we let the cells in the middle of the
region proliferate at a constant rate and we terminate the simulation
when $\Ncells = 1000$. The average in eq.~\eqref{eq:dave} is appropriately
modified to account for any doubly occupied voxels. To get a more
realistic contact pattern a hexagonal grid was used; hence each voxel
has six neighbours.

\begin{figure}[H]
  \centering
  \includegraphics[clip=true,trim=1cm 0.7cm 1cm 0.7cm]{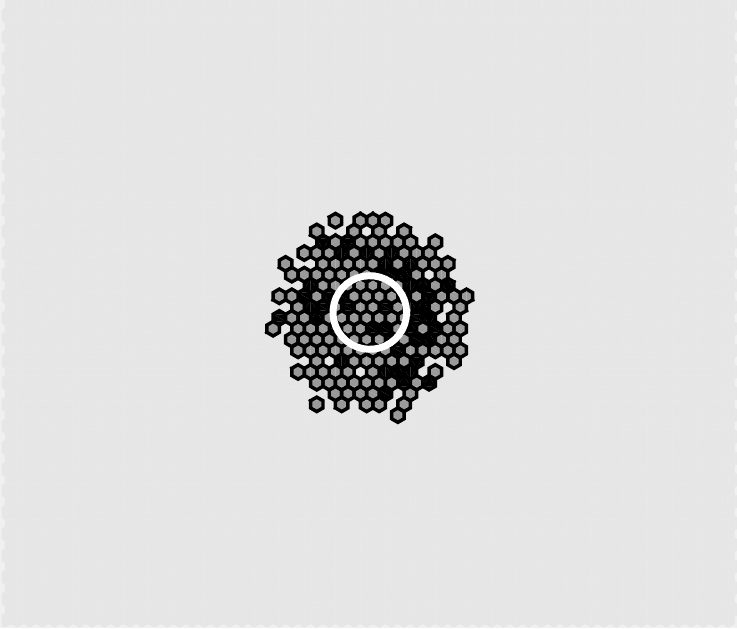}
  \includegraphics[clip=true,trim=1cm 0.7cm 1cm 0.7cm]{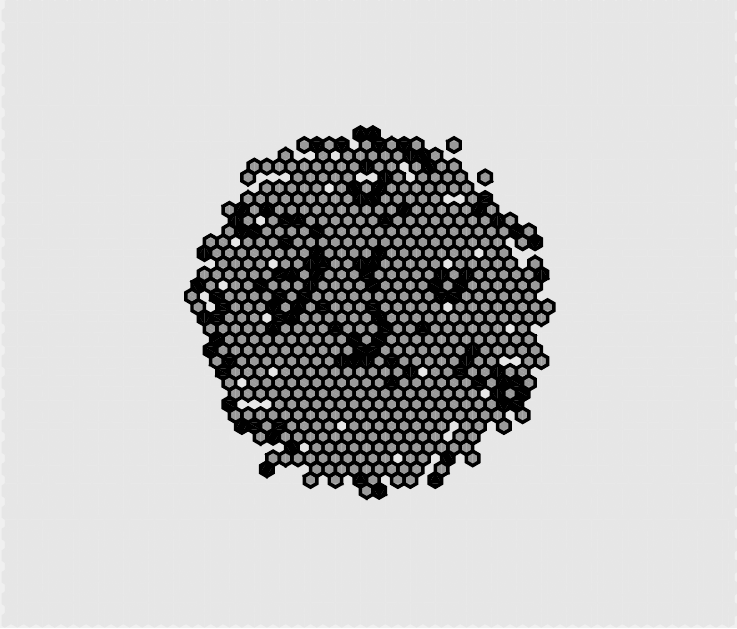}\vspace{0.1cm}
  \includegraphics[clip=true,trim=1cm 0.7cm 1cm 0.7cm]{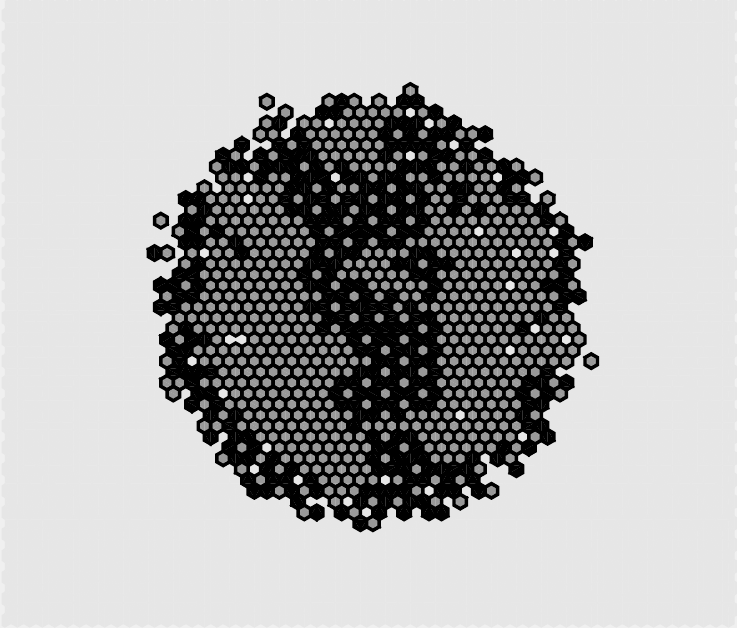}
  \includegraphics[clip=true,trim=1cm 0.7cm 1cm 0.7cm]{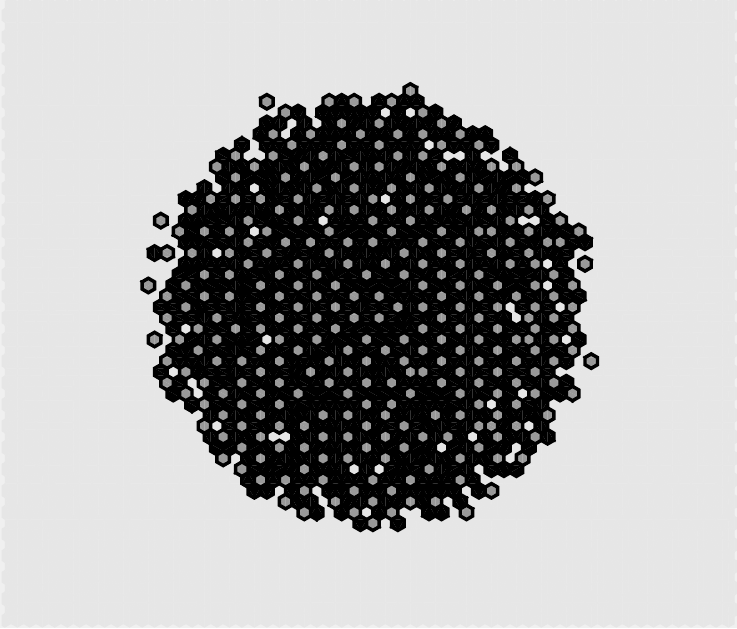}
  \caption{Coupling cellular signalling in continuous time: the
    dynamics of Delta-Notch when cells proliferate. \textit{Black}
    cells indicate high Notch ($n_i > 0.5$) and \textit{grey} cells
    indicate low Notch ($n_i \le 0.5$). \textit{Top left:} initially
    cells are placed in a circular region (demarcated by the white
    circle) and all cells in this region are allowed to proliferate at
    a constant rate. \textit{Top right:} the population thus grows and
    a Delta-Notch signalling model is simulated
    concurrently. \textit{Bottom left/right:} the Delta-Notch model
    equilibrates after the cellular growth process has stopped. The
    times for these snapshots are $t = [5,40,70,100]$ units of time at
    unit proliferation rate, and the model was simulated using the
    Delta-Notch parameters described in the text.}
  \label{fig:delta_notch}
\end{figure}

\begin{figure}[htp]
  \centering
  \includegraphics[clip=true,trim=1cm 0.7cm 1cm 0.7cm]{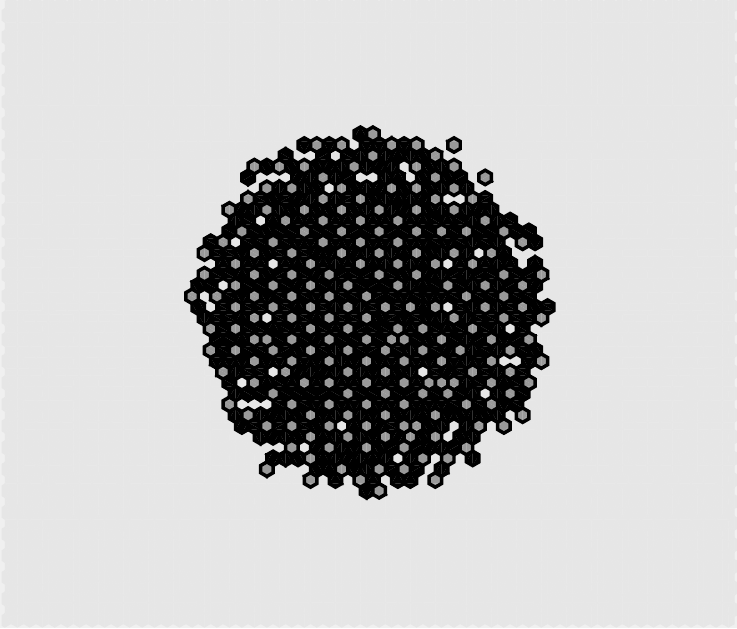}
  \includegraphics[clip=true,trim=1cm 0.7cm 1cm 0.7cm]{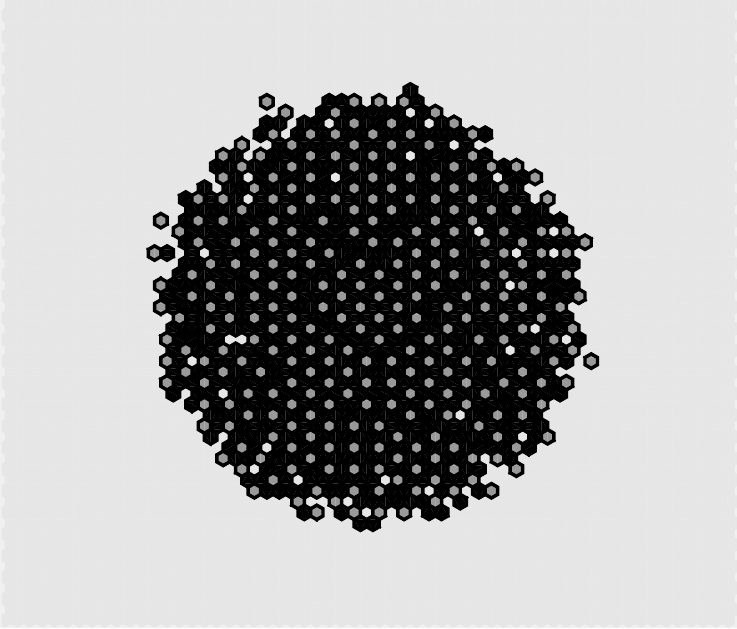}
  \caption{Delta-Notch dynamics simulated anew, but on a faster
    time-scale. Here the process is essentially in continuous
    equilibrium as the tissue grows. These two frames correspond to the
    top right and bottom left frames of Figure~\ref{fig:delta_notch}.}
  \label{fig:delta_notch_fast}
\end{figure}

In Figure~\ref{fig:delta_notch} a typical time-series of this model is
summarised. Here the time-scale of the Delta-Notch dynamics,
eq.~\eqref{eq:delta_notch_ODE}, is on a par with that of the
proliferation process, so that their dynamics equilibrates after the
tissue stops growing. A second simulation is summarised in
Figure~\ref{fig:delta_notch_fast}, where the right-hand side of
eq.~\eqref{eq:delta_notch_ODE} has been scaled by a factor of 50 so
that the Delta-Notch model is in quasi-steady-state as the tissue
grows. The significant differences between the patterns could
potentially be used, together with experimental data, to better
understand how the timescales for patterning compare with those of
tissue growth.

\subsection{Non-trivial dynamics emerges from the combination of grid-based physics and local behaviour}
\label{subsec:dynamics}

Finally, we utilise our proposed framework to build a model for
avascular tumour growth. It is known that an important factor in the
growth of tumours is the availability of oxygen to the tumour. In the
case of an avascular tumour there is no direct supply of oxygen to the
tumour itself, but instead oxygen in the surrounding tissue diffuses
into the tumour. We consider the tumour to be a growing population of
cells that can potentially proliferate and/or die over the course of
the simulation. We extend the occupancy of voxels such that
$u_i \in \{-1,0,1,2\}$, where $u_i=-1$ corresponds to a dead cell
occupying a voxel. The cells at the boundary of the tumour can consume
oxygen and proliferate if the concentration of oxygen is
sufficient. Further into the tumour, where the concentration of oxygen
is generally lower, the cells continue to consume oxygen but can no
longer proliferate; these are known as quiescent cells. Near the
centre of the tumour the oxygen concentration is low enough that the
cells die and eventually degrade; this region is known as the necrotic
core.

As before we solve the discrete Laplace equation (\ref{eq:dlaplace})
for the pressure in each voxel, with homogeneous Dirichlet boundary
conditions (\ref{eq:hom_dirichlet}) on the boundary $\partial
\Omega_h$, the collection of empty voxels that are adjacent to
occupied voxels. We also have an equation for the oxygen
concentration, $c$, which diffuses through the domain with sources at
the fixed external boundary, $\partial \Omegaext$, thus,
\begin{align}
  \label{eq:oxygen}
  -L c & = -\lambda a(u), \\
  \label{eq:oxygen_BCs}
  c_i & = 1, \; i \in \partial \Omegaext,
\end{align}
where $\lambda$ is the rate of consumption of oxygen for a single cell
and $a(u_i)$ is the number of alive cells ($u_i \in \{0,1,2\}$) in the
$i$th voxel (doubly occupied voxels consume twice as much oxygen and
empty voxels or those containing dead cells consume no oxygen). From
these discretised equations we can calculate all the different rates
for the possible events. These events are as follows. A cell occupying
its own voxel, $u_i=1$, will proliferate at rate $\rhoprol$ if
$c_i > \kappaprol$, where $\kappaprol$ is the minimum oxygen
concentration for proliferation to occur. A living cell will die at a
rate $\rhodeath$ if $c_i < \kappadeath$, where $\kappadeath$ is the
threshold oxygen concentration for cell survival. In the case of cell
death, $u_i=1$ is replaced with $u_i=-1$. A dead cell, $u_i = -1$, can
degrade at a constant rate $\rhodeg$ to free up the voxel ($u_i=0$)
for other cells to move in to it. We also include the rates for cell
movement as shown in eqs.~\eqref{eq:D1}--\eqref{eq:D3}.

\begin{figure}[p]
  \centering
  \includegraphics[scale=0.8]{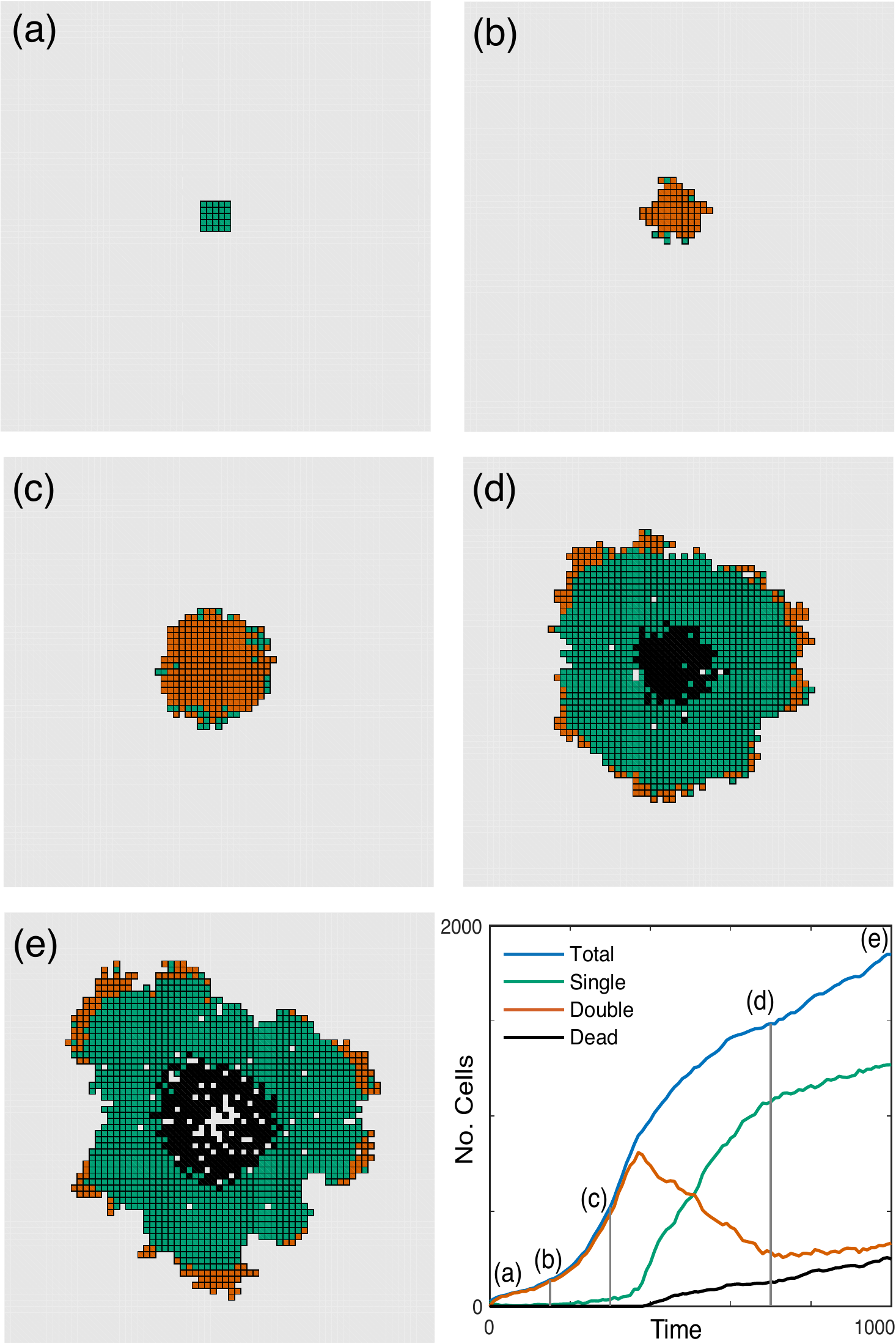}
  \caption{Snapshots of a realisation of the avascular tumour model at
    times $t=[0,150,300,700,1000]$; the model parameters are as given
    in the text. (a): initially cells are placed in a small $5 \times
    5$ grid of singularly occupied voxels (green). (b): cells have
    proliferated exponentially due to an initial abundance of oxygen,
    and most voxels are doubly occupied (red). (c): the tumour grows
    at a quasi-linear rate, mostly voxels are still doubly
    occupied. (d): the lack of oxygen has forced a heterogeneity of
    cell types with a proliferating ring (red), an annulus of
    quiescent cells (green) and a necrotic core (black). The cell
    numbers appear to be plateauing. (e): Asymmetric protrusions form
    on the exterior of the tumour enabling the overall number of cells
    to continue to increase, thereby preventing a complete plateau.}
  \label{fig:tumour}
\end{figure}

\review{In Figure~\ref{fig:tumour} we present snapshots from a realisation of
the model with parameters
\begin{equation*}
D_1=0.01~\text{s}^{-1},
\quad
D_2=25~\text{s}^{-1},
\quad
D_3=0.01~\text{s}^{-1},
\quad
\lambda=0.0015~\text{s}^{-1},
\end{equation*}
\begin{equation*}
\kappaprol=0.65,
\quad
\rhoprol=0.125~\text{s}^{-1},
\quad
\kappadeath=0.55,
\quad
\rhodeath=0.125~\text{s}^{-1},
\quad
\rhodeg=0.01~\text{s}^{-1},
\end{equation*}
that demonstrates
the evolution of the tumour into the classical regions of
proliferating cells, quiescent cells and a necrotic core.} We also
present the evolution of the numbers of each cell type over the
duration of the realisation (Figure~\ref{fig:tumour} \emph{bottom
  right}). The widely-held view in the field is that a limited oxygen
supply leads to a stable, finite-sized
tumour~\cite{Roose:2007:MMO,Hirsch::spheroids,Grimes::RSOC}. Recent
models for avascular tumour growth have been utilised to successfully
predict the growth of tumour cell lines that exhibit sigmoidal growth
curves~\cite{Grimes::sigmoidal}. Sigmoidal curves have three phases:
for an initially small population of cells
(Figure~\ref{fig:tumour}~(a)) in an abundance of oxygen we see
exponential growth (Figure~\ref{fig:tumour}~(b)), which is then
followed by a quasi-linear growth of proliferating cells
(Figure~\ref{fig:tumour}~(c)). Eventually the lack of oxygen causes
three heterogeneous cell types to appear
(Figure~\ref{fig:tumour}~(d)); a proliferating ring, a quiescent
annulus and a necrotic core. It is here where the volume of the
avascular tumour should begin to plateau, as observed in simulations
(Figure~\ref{fig:tumour} \emph{bottom right (d)}).

However, due to a possible lack of surface adhesion in our model we
observe a fourth phase emerge, where asymmetric protrusions appear on
the surface (Figure~\ref{fig:tumour}~(e)) of the tumour which
increases the total uptake of oxygen and hence a further incline in
the total cell number (Figure~\ref{fig:tumour} \emph{bottom
  right}). We sought a parameterisation of our model that did not
exhibit this fourth phase and instead produces a steady-state tumour,
however we could not do so despite a systematic search of parameter
space. We can navigate the parameter space as follows; firstly, select
a user-specified finite size tumour radius and with it solve the
steady-state oxygen equation. Then the parameters
($\lambda,\kappa_{\text{prol}},\kappa_{\text{death}}$) can be chosen
to decide where the proliferating ring and necrotic core should
lie. Next, initialise a population of cells at the required radius
with some proliferation and death rates $\rho_{\text{prol}}$ and
$\rho_{\text{death}}$, and temporarily let $\rho_{\text{deg}}=\infty$
such that dead cells instantly degrade. We can then alter the movement
rates $D_1$, $D_2$ and $D_3$ such that the number of cells
proliferating is roughly equal to the number of cells being instantly
absorbed in the necrotic core; this means that, on average, the size
of the tumour will be constant. Finally, returning to the full
avascular tumour model, we can vary the remaining parameters
($\rho_{\text{prol}},\rho_{\text{death}},\rho_{\text{deg}}$) in an
attempt to build a finite-size tumour. Despite this comprehensive
exploration of the parameter space, which is only possible due to the
efficiency and versatility of the DLCM framework, at best we could
only slow down the growth of the asymmetric protrusions by increasing
$D_2$ to be far greater than $D_1$. \review{That cells invade new
  matrix at a slower rate appears to be a reasonable assumption here,
  but the exact separation in scales between $D_1$ and $D_2$ will need
  to be investigated on a case-by-case basis using cell trajectory
  data collected from relevant experiments.} Thus the current set of
modelling assumptions, which only use passive physics, are not capable
of developing the avascular tumour spheroids seen in experiments. Our
results \review{are in line with previous works
  \cite{Jagiella:2016,Dormann:2002} which both conclude that
  additional mechanisms are necessary in order for tumour growth to be
  stabilised. The suggested mechanisms involve the necrotic cells
  secreting chemicals that can either act as a growth inhibitor for
  healthy cells to become quiescent \cite{Jagiella:2016}, or as a
  chemotactic signalling chemical to attract tumour cells to migrate
  back towards the necrotic core \cite{Dormann:2002}. The inclusion of
  such mechanisms fits naturally into our proposed DLCM framework and
  will be implemented in future, more focussed studies.}


\section{Discussion}
\label{sec:discussion}

\review{The goal of this work was to provide an efficient, hybrid and
  multiscale computational framework for modelling populations of
  cells at the tissue scale. Our framework can harness either a
  regular or unstructured lattice, as required by the biological
  problem under consideration, and it draws upon a constitutive,
  mechanistic description of cellular biomechanics to drive expansion
  and/or retraction of the cell population as cells move and undergo
  proliferation and death.} By assuming that the timescale on which
the tissue relaxes to mechanical equilibrium is much shorter than that
of other mechanical processes, we are able to assume that the
``cellular pressure'' within the population is governed by the Laplace
equation with specified boundary conditions and source terms. We then
use the pressure gradient within the tissue to derive rate equations
for the movement of cells within the population. This enables us to
propagate the model in continuous time, using an event-driven
algorithm such as that originally proposed by
Gillespie~\cite{Gillespie:1976:GMN}. A significant advantage of our
approach in this regard is that both discrete and continuous models of
biochemical signalling can easily and flexibly be incorporated into
the framework. This enables the user to develop hybrid and multiscale
models that include constitutive descriptions of cellular
biomechanics, inter- and intra-cellular signalling in an efficient and
scalable framework. \review{Our approach is flexible, computationally
  efficient and provides a consistent description of cellular
  mechanics. However, it does not, in the form presented here, allow
  for detailed tracking of individual cells, or a resolution of their
  shape, over time; other cell-based modelling approaches may be more
  suitable if such a spatial resolution is
  required~\cite{Van-Liedekerke:2015:STM,Fletcher:2017:MCM}.}

We have demonstrated the utility of our approach using four
examples. First, we demonstrated that, as expected, a simple colony of
proliferating cells relaxes isotropically. Second, we demonstrated how
to integrate signals from the local microenvironment into cell
movement laws, using the migration of cells within a neuronal explant
exposed to a gradient in the secreted protein Slit as an example. Our
third example showed that it is easy to integrate cell-cell signalling
models into the framework, using the delta-notch lateral inhibition
model as a test case. Finally, we developed a model for avascular
tumour growth, wherein cell proliferation and death is controlled by
the local oxygen concentration, and the cell population generates an
oxygen profile within the tumour as the cells consume oxygen.

In summary, the real advance provided by our approach is the ability
to quickly and efficiently simulate the behaviour of large populations
of cells, in both two and three spatial dimensions. Our model is both
hybrid and multiscale in nature; it can incorporate both continuum and
discrete models of biochemical signalling both within and between
cells. For the purpose of experimenting we have developed a serial
Matlab implementation of the model that employs a direct factorisation
of the discrete Laplace operator and, as such, scales conveniently to
about 20,000 cells. Using a compiled language and parallelisation one
could likely improve upon this figure to some extent. However real
savings in computing time will require optimal Laplace solvers that
utilise, for example, algebraic or geometric multigrid
techniques~\cite{reviewAMG,bookGMG} that can take advantage of the
incremental nature of the computational process when the domain
evolves slowly. We leave this aspect of the implementation for future
work.

In the modern era of biology, where quantitative data describing the
evolution of cell populations and tissues can be collected with
relative ease, we are now in a position to test and validate
experimentally generated hypotheses using biologically realistic
mechanistic models. However, these models need to be calibrated
against the available data, and the sensitivity of model predictions
to changes in model parameters needs to be explored. All of this
requires repeated simulation of multiscale and hybrid cell-based
models, the computational demands of which have to-date provided a
barrier to significant progress. The DLCM method we outline here
results in significant advances in our ability to efficiently simulate
hybrid and multiscale cell-based models in two and three spatial
dimensions, and therefore provides a very real opportunity to test and
validate mechanistic models using quantitative data.

\subsection{Availability and reproducibility}
\label{subsec:reproducibility}

The computational results can be reproduced within release 1.4 of the
URDME open-source simulation framework, available for download at
\url{www.urdme.org}.

\subsection*{Acknowledgement}

REB would like to thank the Royal Society for Wolfson Research Merit
Award and the Leverhulme Trust for a Leverhulme Research Fellowship. DBW would like to thank the UK's Engineering and Physical
Sciences Research Council (EPSRC) for funding through a studentship at
the Systems Biology programme of The University of Oxford's Doctoral
Training Centre.


\bibliographystyle{abbrvnat}
\bibliography{bew.bib}

\end{document}